\documentclass[pra,amsmath,amssymb,twocolumn,superscriptaddress]{revtex4}

\usepackage{graphicx}
\usepackage{epstopdf} 
\usepackage{dcolumn} 
\usepackage{bm}
\usepackage{subfigure}
\usepackage{color}

\newcommand{\ket}[1]{| #1 \rangle}
\newcommand{\Ham}{{\cal H}}
\newcommand{\Id}{\mathbb I}

\begin{document}

\title{Quantum discord in a spin system with symmetry breaking}

\author{Bruno Tomasello}
\affiliation{MATIS-INFM-CNR $\&$ Dipartimento di Fisica e Astronomia, 95123 Catania, Italy}
\affiliation{SEPnet and Hubbard Theory Consortium, University of Kent, Canterbury CT2 7NH, U.K.}
\affiliation{ISIS facility, Rutherford Appleton Laboratory, Harwell Oxford Campus, Didcot OX11 0QX, U.K.}

\author{Davide Rossini}
\affiliation{NEST, Scuola Normale Superiore $\&$ Istituto Nanoscienze-CNR, Piazza dei Cavalieri 7, I-56126 Pisa, Italy}

\author{Alioscia Hamma}
\affiliation{Perimeter Institute for Theoretical Physics, 31 Caroline St. N,Waterloo ON, N2L 2Y5, Canada}

\author{Luigi Amico}
\affiliation{MATIS-INFM-CNR $\&$ Dipartimento di Fisica e Astronomia, 95123 Catania, Italy}

\begin{abstract}
  We analyze the \emph{quantum discord} $Q$ throughout the low-temperature phase diagram of the
  quantum XY model in transverse field. We first focus on the $T=0$ order-disorder quantum phase transition 
  both in the symmetric ground state and in the symmetry broken one.
  Besides it, we highlight how $Q$ displays clear anomalies also at a non critical value 
  of the control parameter inside the ordered phase, where the ground state is completely factorized.
  We evidence how the phenomenon is in fact of collective nature and displays universal features. 
  We also study $Q$ at finite temperature. We show that, close to the quantum phase transition, 
  $Q$ exhibits quantum-classical crossover of the system with universal scaling behavior. 
  We evidence a non trivial pattern of thermal correlations resulting from the factorization phenomenon. 
\end{abstract}

\maketitle


\section{Introduction}  \label{sec:intro}

Correlations provide a characterization of many-body systems~\cite{wenbook}.
In the quantum realm both classical and non local quantum correlations 
(like \emph{entanglement}) are relevant.
Although entanglement completely describes quantum correlations for \emph{pure} states, 
it is in general subtler to characterize the pattern of correlations for \emph{mixed} states.
Indeed the quantitative interplay between classical and quantum correlations has been formulated 
only recently with the introduction of the {\it quantum discord}, operatively 
defining quantum correlations in composite systems~\cite{zurek, vedral}.
In this context, intense research activity has been devoted to spin models,
generically displaying non trivial patterns of correlations in different interesting 
regimes~\cite{amico, first_ising, sarandy, topo, maziero, rigolin_temp, recent}.

Two peculiarities are relevant here: on one hand, the order-disorder 
Quantum Phase Transitions (QPT) occurring at zero temperature
as long as the control parameter $h$ is tuned across a critical value $h_c$~\cite{sachdev}.
It is worth noting that the quantum order arises because superselection rules lead to a symmetry breaking~\cite{coleman}. 
Besides QPT, spin systems may exhibit a further remarkable phenomenon occurring at $h=h_f$ 
located within the ordered symmetry-broken phase, where the ground state is exactly 
factorized~\cite{factorization, giampaolo}, and therefore correlations are exclusively classical. 
Such factorization consists in a quantum transition exclusively for entanglement~\cite{fubini}
and is rigorously not accompanied by any change of symmetry. 

In this article we provide the details of the analysis of the quantum discord arising 
in the quantum $XY$ spin-$1/2$ system, both at zero and at finite temperature~\cite{tomasello}.
In particular we consider the ground state with broken symmetry.
We analyze the quantum discord both at the QPT and close to $h_f$.
We evidence how the quantum criticality affects the quantum discord 
at low-temperature, allowing to define the quantum cross-over phenomena in an operative way.
Close to the factorizing point, the finite-temperature quantum discord displays characteristic traits.

The structure of this paper is as follows. 
In Sec.~\ref{sec:overview} we give an overview on the notions of quantum and classical correlations 
in a general quantum system. In Sec.~\ref{sec:XYmodel} we introduce a many body system 
that is suitable for our type of analysis: the $XY$ model in an external transverse field.
We also present some features related to its physics and to the way 
in which correlations are evaluated (Sec.~\ref{sec:corr}).
The analysis of our results is carried out in the subsequent sections,
first for the zero-temperature case (Sec.~\ref{sec:QDiscord_T0})
and then when the temperature is switched on (Sec.~\ref{sec:QDiscord_Tfin}). 
Finally in Sec.~\ref{sec:concl} we draw our conclusions.

\section{Quantum, Classical and Total Correlations}  \label{sec:overview}

In a bipartite quantum system $ AB $ the total amount of correlations between $ A $ and $ B $ 
is given by the mutual information
\begin{equation}
  I(A:B) \equiv {\cal S}(\hat{\rho}^A) + {\cal S}(\hat{\rho}^B) - {\cal S}(\hat{\rho}^{AB}) \,,
  \label{mutual_def}
\end{equation}
where ${\cal S}(\hat{\rho}) = - {\rm Tr} [\hat{\rho} \log_2 \hat{\rho}]$ is the von Neumann entropy.
In classical information, using the \emph{Bayes rule}, an equivalent formulation of the mutual information is
\begin{equation}
  {\cal J}(A:B) \equiv {\cal S}(A) - {\cal S}(A|B) \, ,
  \label{mutual_def_2}
\end{equation}
where the conditional entropy $ {\cal S}(A|B)= {\cal S}(AB) - {\cal S}(B) $ quantifies 
the ignorance on part A once a measurement on B is performed.
In the quantum realm, a measurement generally perturbs the system and part 
of the information itself is lost. Therefore, when we consider a quantum composite system, 
Eq.~\eqref{mutual_def_2} differs from Eq.~\eqref{mutual_def}.
This \emph{difference} allows us to estimate the relative role of quantum and classical
correlations in composite systems~\cite{zurek}.

As a matter of fact, if we describe a measurement on part $ B $ by a set of projectors $\{\hat{B}_k\}$, then
\begin{equation}
  \hat{\rho}^{AB}_{(k)} = \frac{1}{p_k} (\hat{\Id}_A \otimes \hat{B}_k) \, \hat{\rho}^{AB} \, (\hat{\Id}_A \otimes \hat{B}_k)
  \label{rho_cond}
\end{equation}
is the composite state conditioned to the $k$-th outcome with probability 
$ p_k =  \text{Tr}[(\hat{\Id}_A \otimes \hat{B}_k) \, \hat{\rho}^{AB} \, (\hat{\Id}_A \otimes \hat{B}_k)] $.
This conditioned state is the key ingredient which allows to distinguish between classical and quantum correlations:
in fact it generally differs from the pre-measurement state $ \hat{\rho}^{AB} $, as well as
the mutual information differs from classical correlations. 
A reasonable definition of classical correlations $C$ then comes by
finding the set of measurements on $\{\hat{B}_k\}$ that disturbs the least the part $A$, 
i.e. by maximizing~\cite{vedral,zurek}
\begin{equation}
  C(\hat{\rho}^{AB}) \equiv 
  \max_{\{\hat{B}_k\}} \big[ {\cal S}(\hat{\rho}^A) - {\cal S}(\hat{\rho}^{AB} | \{\hat{B}_k\}) \big] \,.
  \label{class}
\end{equation}
The difference between mutual information and classical correlations defines the so called \emph{quantum discord}:
\begin{equation}
  Q(\hat{\rho}^{AB}) \equiv  {\cal I} (A:B) - C(\hat{\rho}^{AB}) \,.
  \label{disc_def}
\end{equation}
In the estimate of quantum correlations between subsystems of a bipartite system, 
entanglement has been playing a leading role, in particular about the relevance 
of correlations in many body systems.
However in general quantum discord differs from entanglement:
for example, even if they are the same for pure states, they can display 
a very different behavior in mixed states.

\section{The $XY$ Model in Transverse Field}  \label{sec:XYmodel}

Hereafter we will focus on an interacting pair of spins $1/2$ in the anti-ferromagnetic $XY$ chain 
with transverse field $ h $. The Hamiltonian of the model
\begin{equation}
  \hat{\Ham} = -\sum_j \left(
  \frac{1+\gamma}{2}\hat{\sigma}_{j}^x \hat{\sigma}_{j+1}^x +
  \frac{1-\gamma}{2} \hat{\sigma}_{j}^y \hat{\sigma}_{j+1}^y +
  h \hat{\sigma}_{j}^z\right)
  \label{eq:XYmodel}
\end{equation}
describes the competition between two parts: the nearest neighbor anisotropic interaction 
on the $ xy $ plane, with an anisotropy tuned by varying $\gamma \in \left( 0, 1 \right]$, 
and the coupling with external magnetic field $ h $ along the $ z $ direction
(throughout this paper we set $\hbar = k_B = 1$). 
Using a set of successive transformations (Jordan-Wigner,  Bogoliubov, Fourier~\cite{lieb}), 
the Pauli matrices operators $\hat{\sigma}^\alpha_j$ ($\alpha = x,y,z$) on sites $j$
can be expressed in terms of operators such that the Hamiltonian takes the diagonal form 
\begin{equation}
  \hat{\Ham} = -\sum_k   \Lambda_k \eta^{\dag}_k \eta_k + {\rm const} \,.
\end{equation}
The system is thus described as a gas of non interacting fermionic quasiparticles, 
where $ \eta^{\dag}_k$ ($ \eta_k  $) is the creation (annihilation) operator of a fermion with momentum $ k $.
Furthermore, the Jordan-Wigner transformation allows
an \emph{analytic} expression for the correlation functions 
$g_{\alpha \alpha}(r) = \langle \hat{\sigma}^\alpha_j \hat{\sigma}^\alpha_{j+r} \rangle$ 
of any two spins in the chain far $r$ sites from each other~\cite{mcoy}. 

The exact solution of the $ XY $ model encouraged a plethora of studies concerning 
its critical phenomena~\cite{mcoy,pfeuty}. In particular, during the 
last decade, new insight has been made in the description of the physics of the system
through the analysis of quantum correlations (i.e. \emph{entanglement})~\cite{amico}.

\subsection{The Phase Diagram}

The phase diagram of the $ XY $ model is marked by \emph{two} peculiar values 
of the applied field $ h $~\cite{amico, sachdev, factorization}.
For $\gamma \in \left( 0, 1 \right]$ the system displays a zero-temperature ($T=0$) continuum 
QPT at $h_c = 1$, of the Ising universality class 
with critical indices $\nu=z=1$, $\beta=1/8$~\cite{sachdev}.
In fact, for strong enough external fields ($ h\gg h_c $) all the spins tend
to be aligned along the $ z $ direction, while the opposite 
limit ($ h\ll h_c $) gives rise to a spontaneous magnetization (with $ Z_2$ symmetry broken)
along a direction on the $ xy $ plane, that is $ \gamma $-dependent.
At zero temperature, on the left side $ h < h_c $ of the phase diagram the system 
is an ordered ferromagnet and the $ Z_2 $ symmetry is broken, 
while on the right side $ h > h_c $ quantum fluctuations 
lead to a disordered phase where the system is a quantum paramagnet.
At finite temperature $T>0$ the physics of the whole system 
is still affected by the QCP $h=h_c$ at zero temperature.
A $V$-shaped diagram in the $h-T$ plane emerges, characterized by 
the straight lines $ T= | h-h_c | $ that mark the crossover region 
between the so called \emph{quantum critical region} ($ T> | h-h_c | $)
and the \emph{quasi-classical} regions surrounding it~\cite{sachdev}.

Besides the QCP, there is another value of the transverse field that characterizes 
the zero-temperature phase diagram of the $ XY $ model.
Indeed it has been found that, at zero temperature and at a certain anisotropy $ \gamma $,
for $ h= h_f = \sqrt{1-\gamma^2} $ the ground state is exactly factorized~\cite{factorization}: 
\begin{equation}
  \label{eq_FactorState}
  \ket{\Psi_{GS}^{\gamma}} = \prod_j \ket{\psi^{\gamma}_j} \,,
\end{equation}
$\ket{\psi^{\gamma}_j}$ being a normalized single-site pure state.
Therefore it emerges that, even though the system is in a phase with very strong quantum correlations, 
there is a ``critical'' set of values $ h_f(\gamma) $ where the state is completely classical.
This strange occurrence, regarded as a paradox in the first place~\cite{factorization},
seems to be strongly connected with the reshuffling of correlations among the system.
In fact, a deep analysis on the behavior of entanglement 
has remarkably shed new light on the relevant physics involved on $ h_f $~\cite{giampaolo}.
In particular it has been shown that, tuning the external field from $ h<h_f $
to $ h>h_f $, the entanglement pattern swaps from parallel to anti-parallel~\cite{fubini}.
Furthermore it has been observed that at zero temperature
the bipartite entanglement has a logarithmically divergent range at $ h_f $,
together with the fact that at finite temperature there is a whole region 
fanning out from $ h_f $ where no pairwise entanglement survives~\cite{amico06}.\\
This strongly suggests that, along these critical values of field and temperature, 
the behavior of entanglement, and in general of correlations, plays a pivotal role 
in the physics involved and hence in our understanding of it.
In particular it seems that the interplay of correlations when 
the field is tuned across $ h_f $ is the only accessible way, so far, 
to tackle the puzzling physics that leads to the factorized state of Eq.~\eqref{eq_FactorState}.
Here we show that the quantum discord allows a fine structure of the phase diagram around $ h_c $,
and, most remarkably, it displays a non trivial scaling law at the factorization field $ h_f $.

\section{Classical and Quantum Correlations in the $XY$ model}  \label{sec:corr}

In order to compute $ Q_r $ between any two spins $ A $ and $ B $
at distance $ r $ along the chain, the key ingredients are 
the single-site density matrices $ \hat{\rho}^A,\hat{\rho}^B $ and
the two-site density matrix of the composite subsystem $ \hat{\rho}^{AB} $ [see Eq.\eqref{disc_def}].
Due to translational invariance along the chain, single-site density matrices 
are the same for any spin and they are given by
\begin{equation}
  \label{1site_rho}
  \hat{\rho}^A = \hat{\rho}^B = \frac{1}{2} \left(\begin{array}{cc} 1 + g_z & g_x \\ g_x & 1 - g_z \end{array} \right) \, ,
\end{equation}
where $g_\alpha \equiv \langle \hat{\sigma}_\alpha \rangle$ are the local expectation values of the magnetization
along the three different axes.
On the other hand, the expression of $ \hat{\rho}^{AB} $ may be cumbersome.
In fact, the general two-site reduced density matrix for a Hamiltonian model
with global phase flip symmetry has the following form~\cite{palacios}:
\begin{equation}
  \label{rhoAB}
  \hat{\rho}_r= \frac{1}{4}
  \left(\begin{array}{cccc}\text{A} & \text{a} & \text{a} & \text{F} \\ \text{a} & \text{B} & \text{C} & \text{b} \\ \text{a} & \text{C} & \text{B} & \text{b} \\ \text{F} & \text{b} & \text{b} & \text{D}\end{array}\right)
\end{equation}
in the basis $ \{ \ket{00} , \ket{01} , \ket{10} , \ket{11} \} $, 
where $ \ket{0} $ and $ \ket{1} $ are eigenstates of $ \hat{\sigma}^z $
(because of translational invariance, this density matrix depends only on the distance $ r $ 
between the two spins: $ \hat{\rho}_r \equiv \hat{\rho}^{AB}$). 
The various entries in Eq.~\eqref{rhoAB} are related to the two-point correlators 
$g_{\alpha \beta}(r) \equiv \langle \hat{\sigma}^\alpha_j \hat{\sigma}^ \beta_{j+r} \rangle$ and 
to the local magnetizations, according to the following:
\begin{equation}
  \label{rho_elements_parity}
  \begin{split}
    \text{A}& = 1+g_{z}+g_{zz} , \\
    \text{D}& = 1-g_{z}+g_{zz} , \\
    \text{B}& = 1-g_{zz}        \\
    \text{C}& = g_{xx}+g_{yy} ,  \\
    \text{F}& = g_{xx}-g_{yy}
  \end{split}
\end{equation}
express the parity coefficients, while
\begin{equation}
  \label{rho_elements_symbr}
  \begin{split}
    \text{a}& = g_{x}+g_{xz},\\
    \text{b}& = g_{x}-g_{xz}
  \end{split}
\end{equation}
explicit the contribution from the symmetry breaking.

As long as the system is in the $ Z_2$-symmetric phase,
matrix elements in ``low case'' are null ($ \text{a} = \text{b} = 0 $).
The symmetry breaking manifest itself in $ \text{a,b} \neq 0 $~\cite{palacios}.
In the former case, the remaining non vanishing ``upper case'' entries in Eq.~\eqref{rho_elements_parity}
can be evaluated analytically~\cite{mcoy}, therefore we could use 
a fully analytical approach to compute the quantum discord 
in the so called \emph{thermal ground state}~\cite{sarandy}.
In this state the system approaches the ground state by lowering the temperature 
towards the limit $ T=0 $; for this reason the symmetry is conserved 
and the state is not in the ``true'' degenerate ground state.
In the latter case, the $ Z_2$ symmetry is lost and beside 
the spontaneous magnetization $ g_x $, also the non trivially computable $g_{xz}(r)$ 
correlation appears~\cite{johnson}. In that case we resort to the numerical 
Density Matrix Renormalization Group (DMRG) method for finite systems with open boundaries~\cite{dmrg}.

Once we have access to the density matrices through the correlation functions, 
we can compute the explicit form of the mutual information and the classical correlation
in order to distill the amount of pure quantum correlations in Eq.~\eqref{disc_def}.
Since the reduced density matrix of the single spin is the 
same for any site, the mutual information is given by:
\begin{equation}
  {\cal I}(A:B) = 2 \, {\cal S}(\hat{\rho}^A)  - \sum_{\nu=0}^{3} \lambda_{\nu} \log_2{\lambda_{\nu}} \,,
  \label{mutinf_XY}
\end{equation}
where $ \lambda_{\nu}(r) $ are eigenvalues of $ \hat{\rho}^{AB}_r $.
If one further specializes to $Z_2$-symmetric states, Eq.~\eqref{mutinf_XY} can be readily evaluated
by using the following two ingredients:
$i)$ the single-site density matrix in Eq.~\eqref{1site_rho} turns out to be diagonal,
therefore its von Neumann entropy is ${\cal S}(\hat{\rho}^A) = \mathcal{S}_{bin} [(1+ g_z)/2]$,
where $ \mathcal{S}_{bin}(p) = -p \log_2 p -(1-p) \log_2 (1-p) $ is the binary Shannon entropy;
$ii)$ in terms of the correlation functions, the eigenvalues of $\hat{\rho}^{AB}_r$ turn out to be~\cite{sarandy}:
\begin{equation}
\begin{split}
  \lambda_{0/1} &= \frac{1}{4} \Big(1 + g_{zz}  \pm  \sqrt{g^{2}_{z} + (g_{xx} - g_{yy})^{2} } \Big) \\
  \lambda_{2/3} &= \frac{1}{4} \Big(1  -  g_{zz}  \pm  \vert g_{xx} + g_{yy} \vert \Big) \; .
\end{split}
\label{eigen_lambda}
\end{equation}

Once the mutual information~\eqref{mutinf_XY} is known in terms of the correlation functions, 
one needs to perform a suitable measurement on spin $B$ in order 
to compute classical $C(\hat{\rho}_r)$ and eventually quantum $Q(\hat{\rho}_r)$
correlations, as stated in Eq.~\eqref{disc_def}.
Following a procedure similar to Refs.~\cite{first_ising, luo, sarandy},
we use a set of projectors $ \{\hat{B}_k\} $ as local measurements on the spin $ B $.
In particular, working on the computational basis $\{ \ket{0} , \ket{1} \}$
in the Hilbert space $\mathcal{H}^{2}_{B}$ associated to the spin $ B $,
our general set of projectors is 
\begin{equation}
 \{ \hat{B}_{k} = V \hat{\Pi}_{k}V^{\dag} \} \quad,\quad k=0,1
\label{projectors}
\end{equation}
where $ \hat{\Pi}_{k}= | k \rangle \langle k|$ is related to the basis vectors and $V \in U(2)$ 
gives the generalization to any type of projector on $ B $. 
As suggested in Ref.~\cite{sarandy}, $V$ can be parametrized as follows:
\begin{equation}
  V = \begin{pmatrix} \cos \frac{\theta}{2} & \sin \frac{\theta}{2} \, e^{-i\phi} \\[1.1ex]
    \sin \frac{\theta}{2} \, e^{i\phi} & -\cos \frac{\theta}{2}
  \end{pmatrix} \, ,
  \label{parametr_V}
\end{equation}
where $\theta \in[0,\pi]$ and $\phi \in[0,2\pi)$ are respectively the azimuthal and polar axes 
of a qubit over the Bloch sphere in $\mathcal{H}^{2}_{B}$.
After a measurement has been performed on $ B $, the reduced density matrix 
$ \hat{\rho}^{AB} |_{\{\hat{B}_k\}} $ will be in one of the following states:
\begin{equation}
    \hat{\rho}^{AB}_{0/1} = \frac{1}{2} \bigg( \hat{\Id}_{A} + \sum_{\alpha =1}^{3}q_{0/1, \alpha} \hat{\sigma}_{A}^{\alpha} \bigg) \otimes \Big( V \hat{\Pi}_{0/1} V^{\dag} \Big)_{B} \,.
  \label{rho_k}
\end{equation}
This expression gives the explicit dependence of the system $ A $ 
with respect to the projective measurement performed on the spin $ B $. 
The coefficient $ q_{k, \alpha}=q_{k, \alpha}(\theta,\phi) $ in
the expansion depends on the projectors used to perform 
the measure on $ B $ (see Ref.~\cite{sarandy} for the explicit form). 

Finally, the maximization over all possible $\hat{B}_k$ embedded in 
Eq.~\eqref{class} is equivalent to find those values $( \theta_{opt}, \phi_{opt} )$
that \emph{disturb} the least the spin $ A $ when we make a measure on $ B $. 
For $Z_2$-symmetric states we analytically found $ \theta_{opt} = \frac{\pi}{2} , \phi_{opt} = 0$, 
in accordance with Refs.~\cite{sarandy,maziero}.
This eventually leads to a simple closed expression for the classical correlations:
\begin{equation}
C(\hat{\rho}_r) = \mathcal{S}_{bin}(p_1) - \mathcal{S}_{bin}(p_2) \,,
\label{classical_XY}
\end{equation}
where $\mathcal{S}_{bin}(p)$ is the Shannon entropy and
\begin{equation}
  p_{1} = \frac{1+g_z}{2} \,; \qquad
  p_{2} = \frac{1}{2} + \sqrt{ \frac{g_{xx}^{2}+g_z^2}{4}} \, .
  \label{p1_p2}
\end{equation}
By taking the difference between the mutual information~\eqref{mutinf_XY}
and Eq.~\eqref{classical_XY} for the classical correlations,
one gets the following simplified expression for the quantum discord
between any two spins in the $ XY $ chain in transverse field, valid
for $Z_2$-symmetric states~\cite{sarandy}:
\begin{equation}
  Q(\hat{\rho}_r) = \mathcal{S}_{bin}(p_1) + \mathcal{S}_{bin}(p_2) - \sum_{\nu=0}^{3} \lambda_{\nu} \log_2 \lambda_{\nu} \, .
  \label{disc_XY}
\end{equation}

\section{Quantum Discord at T = 0}  \label{sec:QDiscord_T0}

In this section we analyze quantum correlations both in the thermal ground state 
and in the symmetry broken one. In particular we remark differences and similarities 
between them, and highlight the interesting features occurring at the QCP 
and at the factorizing field 
(if not specified, the $xy$ anisotropy is set to $ \gamma = 0.7 $ in every picture).

\begin{figure}[!t]
  \centering
  \includegraphics[width=\columnwidth]{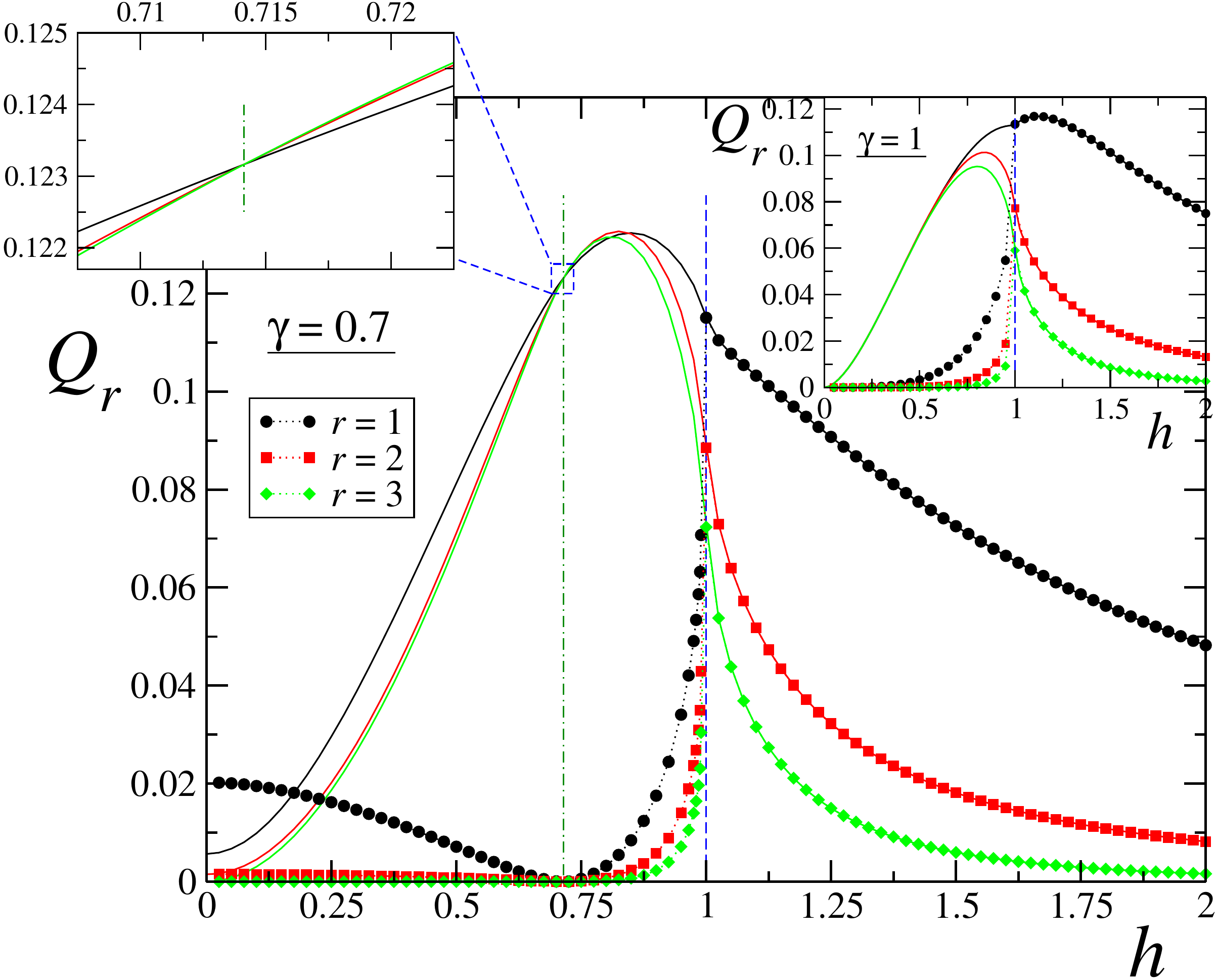}
  \caption{Quantum discord $Q_r(h)$ between two spins at distance $r$ in the $XY$ model 
    at $\gamma = 0.7$ (main plot and left inset) and at $\gamma = 1$ (right inset), 
    as a function of the field $h$.
    Continuous lines are for the thermal ground state, while symbols denote the
    symmetry-broken state obtained by adding a small symmetry-breaking 
    longitudinal field $h_x = 10^{-6}$ and it was computed with DMRG
    in a chain of $L=400$ spins; simulations were performed by keeping $m=500$ states
    and evaluating correlators at the center of the open-bounded chain. 
    For $\gamma=0.7$ and at $h_f \simeq 0.714$, in the symmetric state all the curves 
    for different values of $r$ intersect, while after breaking the symmetry
    $Q_r$ is rigorously zero.
    At the critical point $Q_r$ is non analytic, thus signaling the QPT. 
    In the paramagnetic phase, there is no symmetry breaking affecting $Q_r$.}
  \label{qd}
\end{figure} 

We start by showing the behavior of quantum discord 
at zero temperature, over a wide range of external field values $ h $
centered around the critical value $ h_c $, where the QPT occurs. 
In Fig.~\ref{qd} we plot the quantum discord $Q_r$, for both the $ XY $ model 
(main panel, $ \gamma = 0.7 $) and the Ising model (right inset, $ \gamma = 1 $), 
in the true ground state obtained by means of DMRG simulations (dotted lines with symbols)
and in the thermal ground state evaluated analytically (solid lines without symbols).
In the disordered phase $ h>h_c $ no difference occurs between the two state,
while in the opposite regime $ h<h_c $ two different patterns come out.
In fact, in the latter case the order in the system is really achieved only 
when the symmetry of the system is lost.
We achieve this condition by introducing a tiny longitudinal field ($h_x \ll 1$)
in the DMRG computations that leads to the symmetry breaking
and gives out the true ground state, where quantum correlations
are very small as long as $ h<h_c $. 
It is remarkable that they start to increase once the field is tuned immediately 
upper the factorizing field (where all quantum correlations must vanish), 
eventually reaching a cuspid-like maximum at the QCP.
On the other hand, the quantum discord on the thermal ground state 
is a smooth function with respect to the field. 
In general it depends on the distance $ r $ between the two spins,
but at the factorizing point it gets the same value for any length scale,
as witnessed by the left inset~\cite{ciliberti}.

\begin{figure}[!t]
  \centering
  \includegraphics[width=\columnwidth]{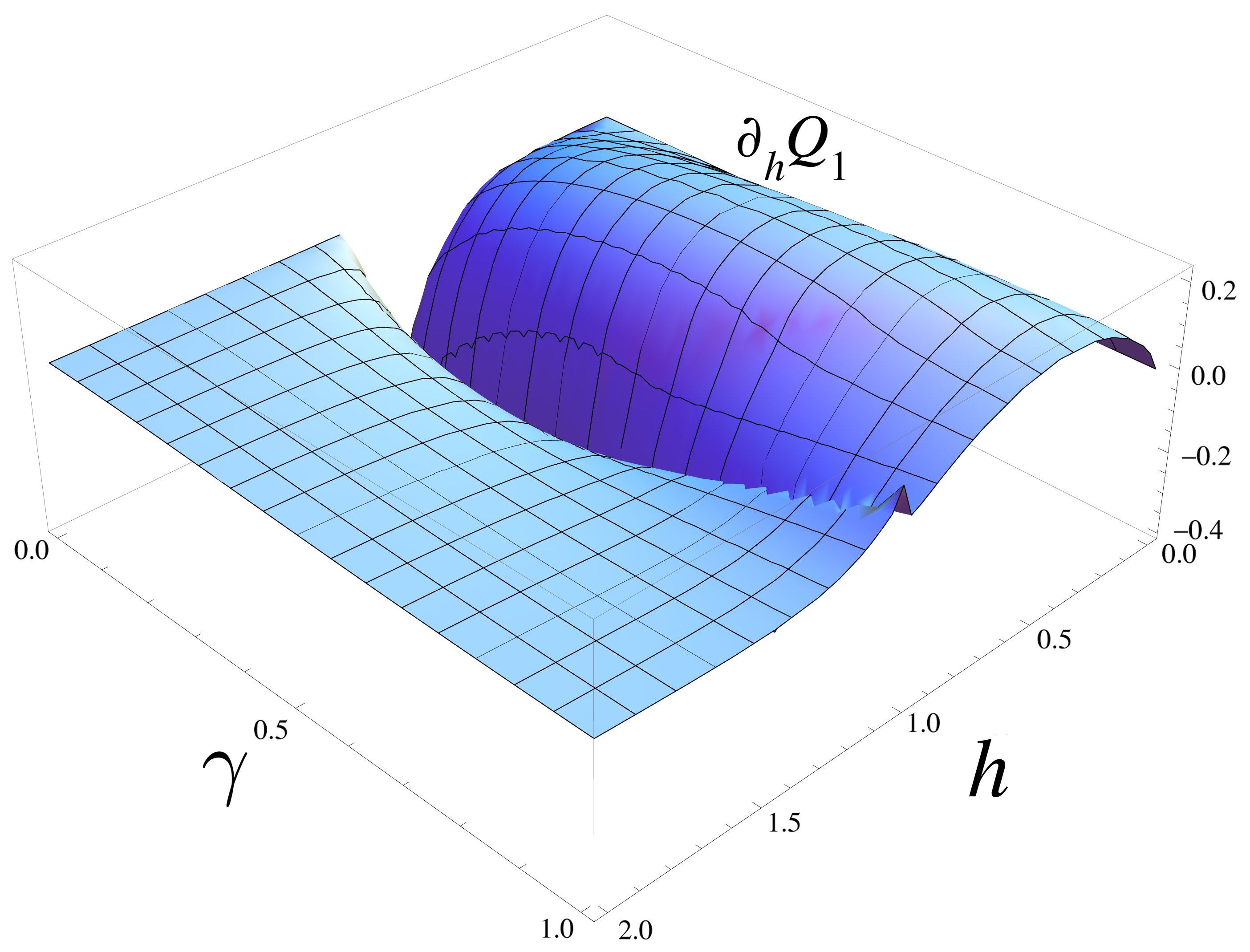}
  \caption{Behavior of $\partial_h Q_r $ in the thermal ground state with respect to the field for any type of anisotropy $ \gamma $.
    Note that for $ \gamma=1 $ at the QCP there is no more divergence but a cuspid.}
  \label{3D_deriv}
\end{figure}

\begin{figure}[!t]
  \centering
  \includegraphics[width=\columnwidth]{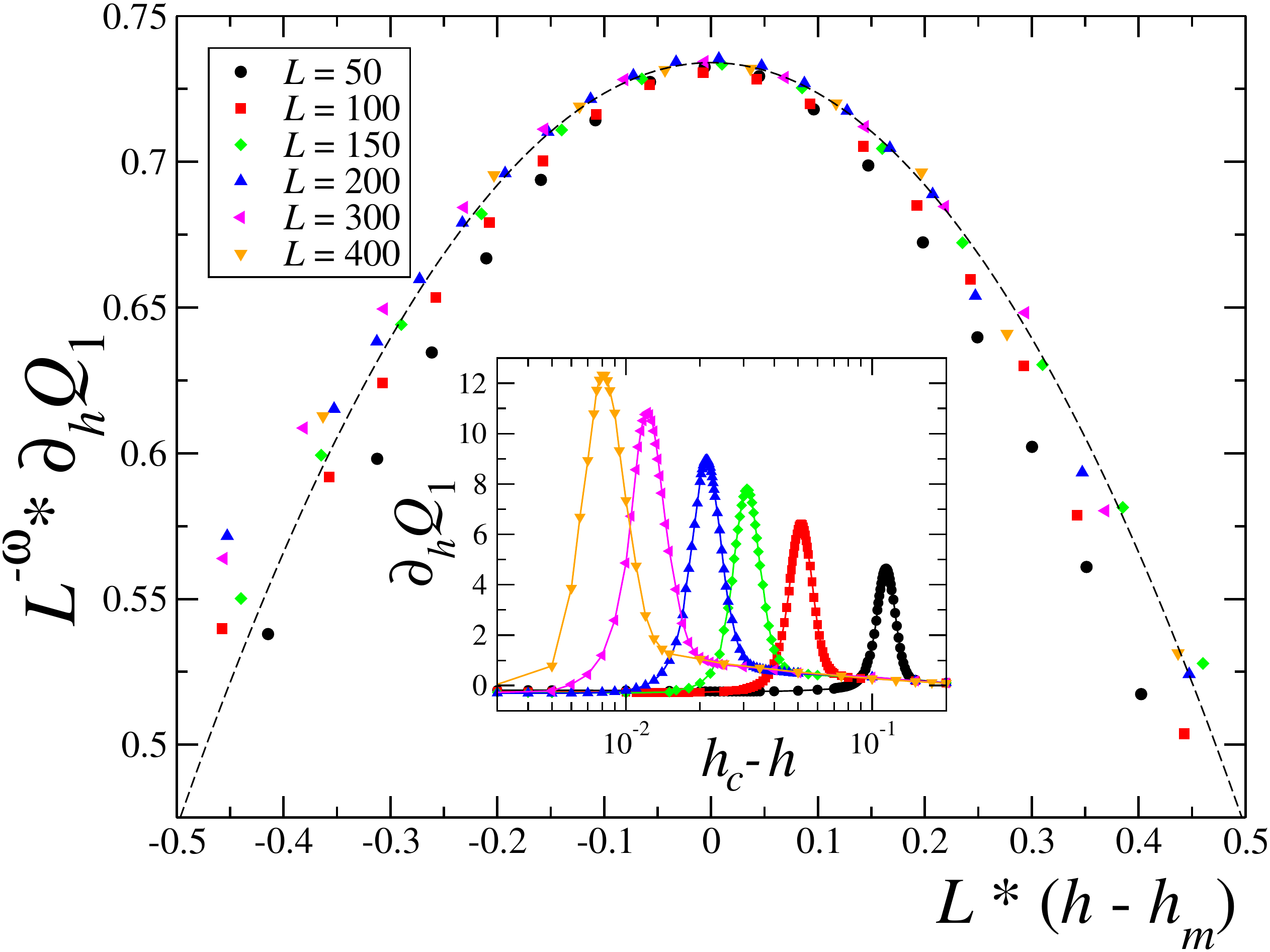}
  \caption{Finite-size scaling of $\partial_h Q_1$ for the symmetry-broken ground state 
    in the $XY$ model, in proximity of the critical point $h_c$. 
    The first derivative of the quantum discord is a function of $L^{-\nu}(h-h_m)$ only, 
    and satisfies the scaling ansatz $\partial_h Q_1 \sim L^\omega \times F[L^{-\nu}(h-h_m)]$,
    where $h_m$ is the renormalized critical point at finite size $L$ and $\omega=0.472$. 
    We found a universal behavior $h_c - h_m \sim L^{-1.28 \pm 0.03}$ with respect to $\gamma$. 
    Inset: raw data of $\partial_h Q_1$ as a function of the transverse field.}
  \label{qcritical}
\end{figure}

To go deeper in the analysis, let us first focus on $ h_c=1 $.
The QPT is in general marked by a divergent derivative of the quantum discord, with respect to the field. 
In particular such divergence is present for any $ \gamma $ in the symmetry-broken state,
while on the thermal ground state it holds as long as  $ \gamma<1 $ (see Fig.~\ref{3D_deriv});
for $ \gamma=1 $, $\partial_h Q_r $ is finite at $ h_c $ although the $\partial^2_h Q_r $ diverges ~\cite{sarandy}.
%
%
This divergence suggests that a scaling analysis at the QCP is feasible. 
In particular in Fig.~\ref{qcritical} we show the finite size scaling $\partial_h Q_{r=1} $
for the symmetry-broken ground state in proximity of $ h_c $.  We found that $ z=\nu=1 $, 
thus meaning that the transition is in the Ising universality class.

\begin{figure}[!t]
  \centering
  \includegraphics[width=\columnwidth]{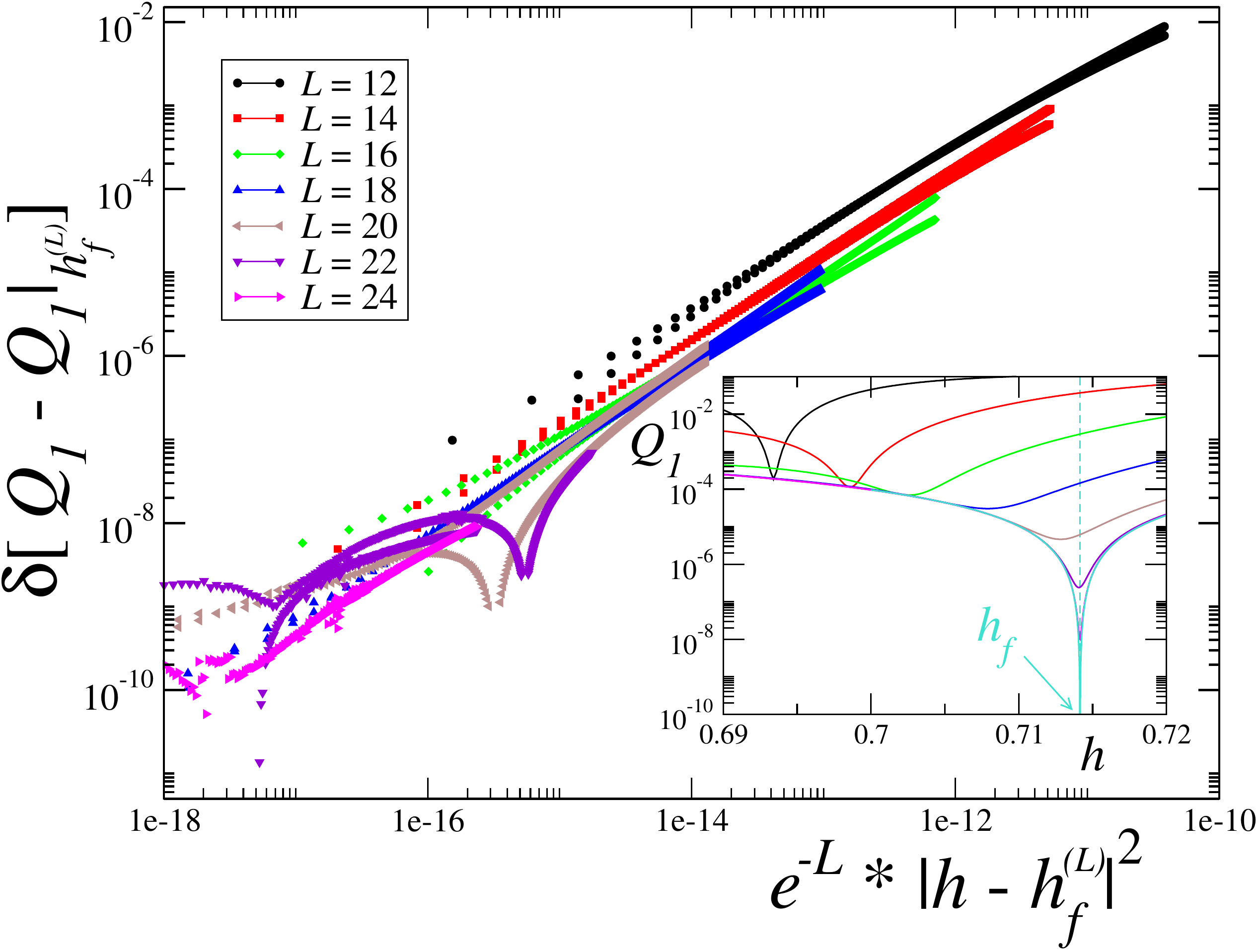}
  \caption{Scaling of $Q_1$ close to the factorizing field:
    we found an exponential convergence to the thermodynamic limit, 
    with a universal behavior according to $e^{-\alpha L}(h-h_f^{(L)})$,
    $\alpha \approx 1$ [$h_f^{(L)}$ denotes the effective factorizing field 
    at size $L$, while $\delta(Q_1) \equiv Q_1^{(L)} - Q_1^{(L \to \infty)}$].
    Due to the fast convergence to the asymptotic value, already at $L \sim 20$ differences 
    with the thermodynamic limit are comparable with DMRG accuracy.
    Inset: raw data of $Q_1$ as a function of $h$. The cyan line is for $L = 30$
    so that, up to numerical precision, the system behaves at the thermodynamic limit.}
  \label{tscaling}
\end{figure}

Turning into the factorizing field $h_f$ we underline that,
for the thermal ground state, it is the only value where the curves with different $ r $,
intersect with each other (see up-left inset in Fig.~\ref{qd})~\cite{ciliberti}.
Besides this, in the broken symmetry state, not only we found that all curves 
vanish in $ h_f $, but we numerically estimated the following dependence of $ Q_r $ close to it:
\begin{equation}
Q_r \sim (h-h_f)^2 \times \big( \frac{1-\gamma}{1+\gamma} \big)^r \,.
\label{Qr@hf}
\end{equation}
Such behavior is consistent with the expression of correlation functions 
close to the factorizing line obtained in Ref.~\cite{baroni}, and here appears 
to incorporate the effect arising from the non vanishing spontaneous magnetization. 
Most remarkably, we found a rather peculiar dependence of $Q_r$ on the system size, 
converging to the asymptotic value $Q_r^{(L \to \infty)}$ with an exponential 
scaling behavior (see Fig.~\ref{tscaling}).
%

\section{Quantum Discord at finite temperature}  \label{sec:QDiscord_Tfin}

Even if both the QCP and the factorizing field occur at $ T=0 $,
they influence the physics of the system even if the temperature is switched on. 
Close to $h_c$, the physics is dictated by the interplay between thermal and quantum fluctuations 
of the order parameter. As we stated before, the cross-over temperature $T_{cross} = |h-h_c|^{z} $ 
fixes the energy scale~\cite{sachdev}.
For $T\ll T_{cross}$ the system is described by a quasi-classical theory,
while inside the ``quantum critical region'' ($T\gg T_{cross}$),
it is impossible to distinguish between quantum and thermal effects.
Here the critical properties arising from the QCP at $ T=0 $ are 
highly dominating the dynamics of the system; as a matter of fact, we expect that 
quantum correlations display some particular pattern as well as they do at $ h_c $.
In fact close to $h_f$ and at small $T$, the bipartite entanglement remains vanishing in a finite 
non linear cone in the $h-T$ plane~\cite{amico,amico06}. 
Thermal states, though, are not separable, and entanglement is present 
in a multipartite form~\cite{toth}. In this regime the bipartite entanglement results to be non monotonous,
and a reentrant swap between parallel and antiparallel entanglement is observed~\cite{amico06}.
At finite temperature, the $ Z_2$ symmetry is preserved 
all over the values of $ h $ (there is no longer a symmetry broken phase).
%
\begin{figure}[!t]
  \centering
  \includegraphics[width=\columnwidth]{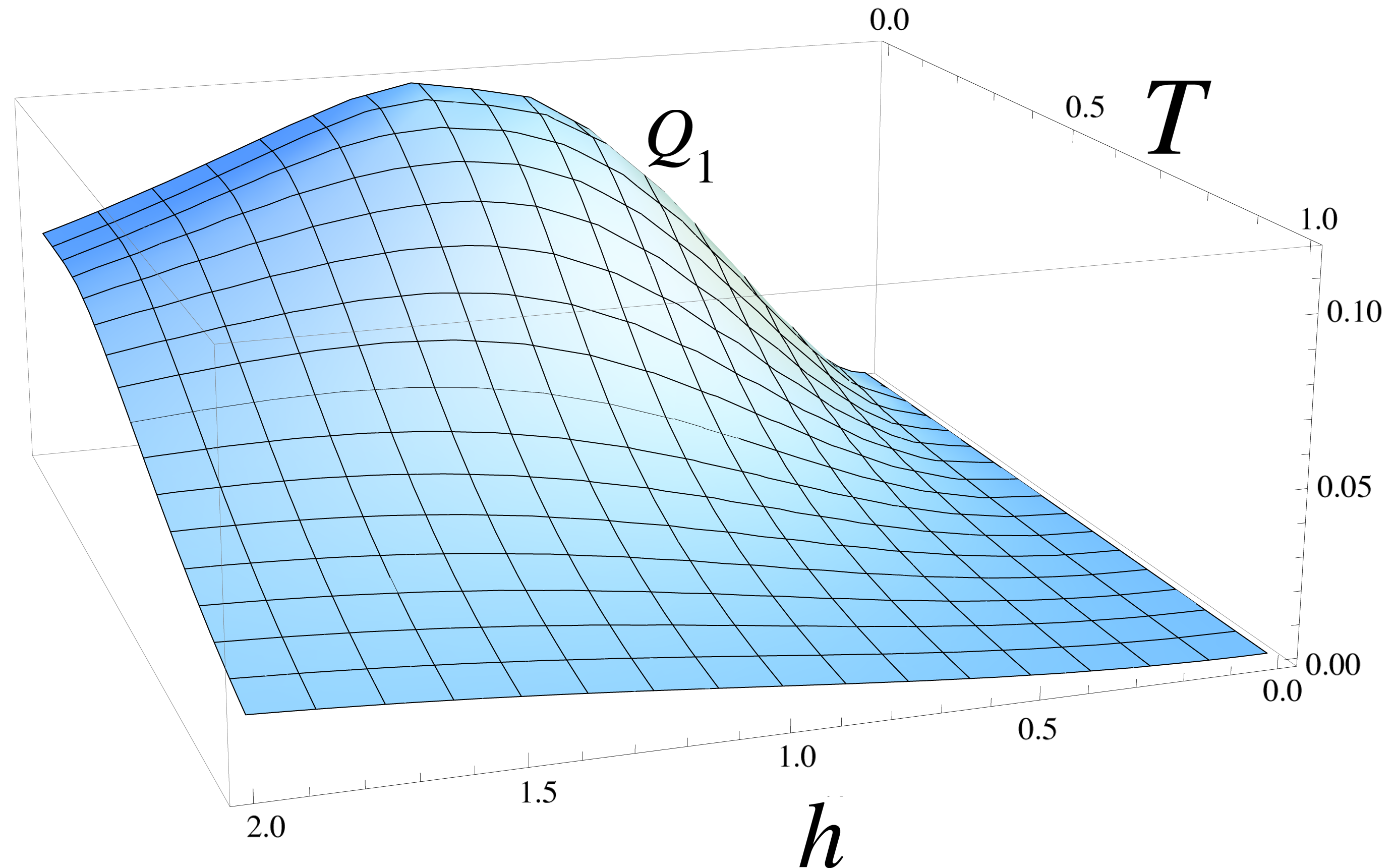}
  \caption{Quantum discord in the thermal state of the Ising model with $ \gamma=1 $, 
    as a smooth function of the temperature $T$ and of the external field $h$.}
  \label{3D_disc_T}
\end{figure} 
%
This means that if the system lies in the ground state at $ T=0 $
(symbols lines in Fig.~\ref{qd}), once the temperature is 
switched on we get a jump of $ Q_r $ all along the phase $ h<h_c $.
After that it behaves as a smooth function decaying with increasing the temperature (Fig.~\ref{3D_disc_T}).
Such discontinuity is also observed in the entanglement, even if in that case 
it is much less pronounced and it occurs only for $h < h_f$~\cite{palacios}.

\begin{figure}[!t]
  \centering
  \includegraphics[width=0.9\columnwidth]{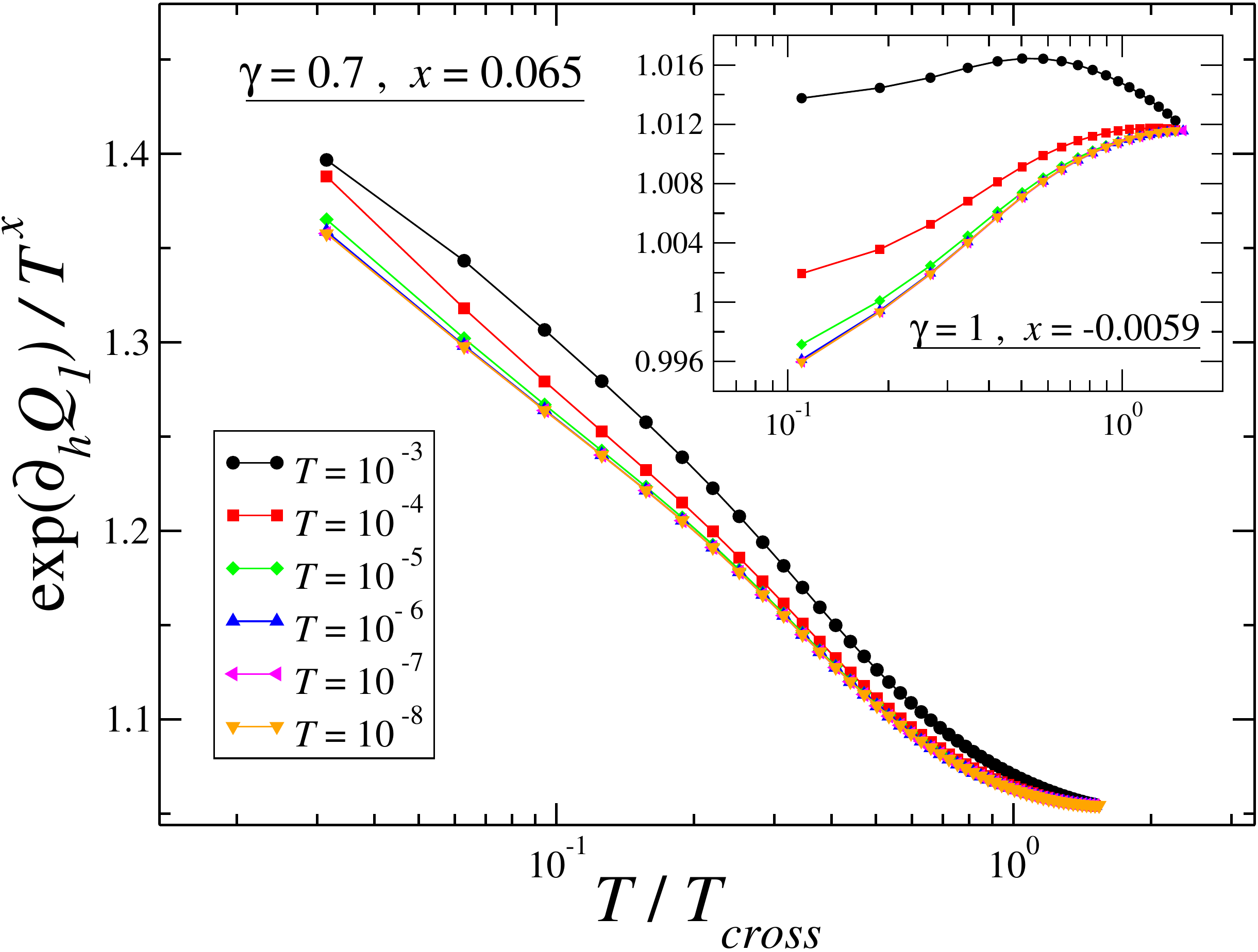}
  \caption{Finite-temperature scaling of the quantum discord for the thermal state 
    close to the critical point. 
    The logarithmic scaling is verified : along the critical line we found   
    $\partial_h Q_1 \vert_{h_c} \sim x \ln(T) + k$, with $x = 0.065$ for $\gamma = 0.7$.
    The scaling function $F$ shows a data collapse close to the critical point.
    Inset: same analysis for the Ising case ($\gamma= 1$); 
    we found an analogous scaling behavior with  $x=-0.0059$.}
  \label{scaling_T}
\end{figure}

\begin{figure}[!t]
  \centering
  \includegraphics[width=0.8\columnwidth]{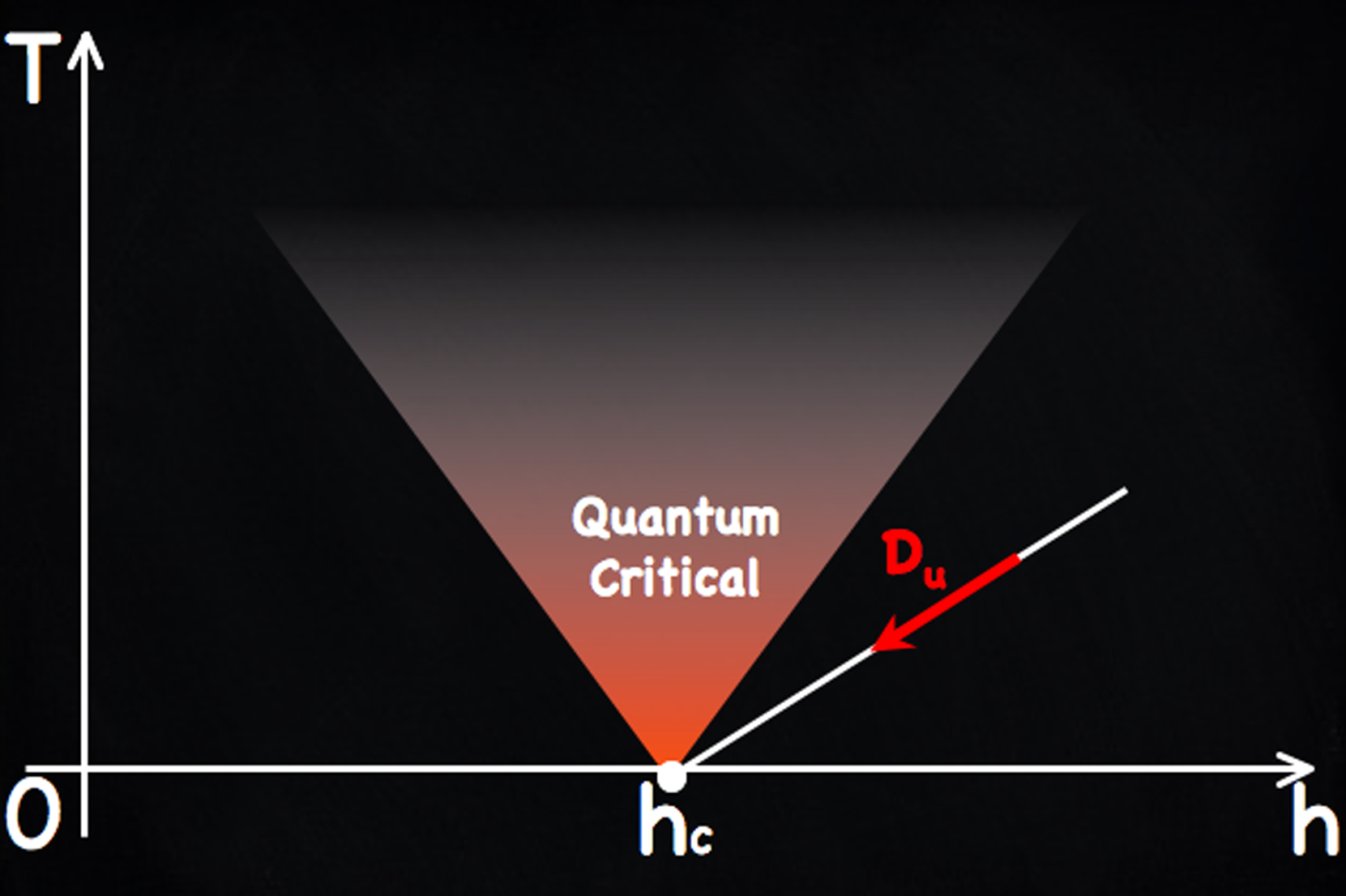}
  \caption{Schematic representation of the directional derivative on the phase diagram in the $h-T$ plane.
    It allows to study how quantities vary along straight lines coming out from 
    the critical point with a slope $\mathbf{u} \equiv (\cos \alpha,\sin \alpha)$. }
  \label{cartoon}
\end{figure}

\begin{figure}[!t]
  \centering
  \includegraphics[width=0.8\columnwidth]{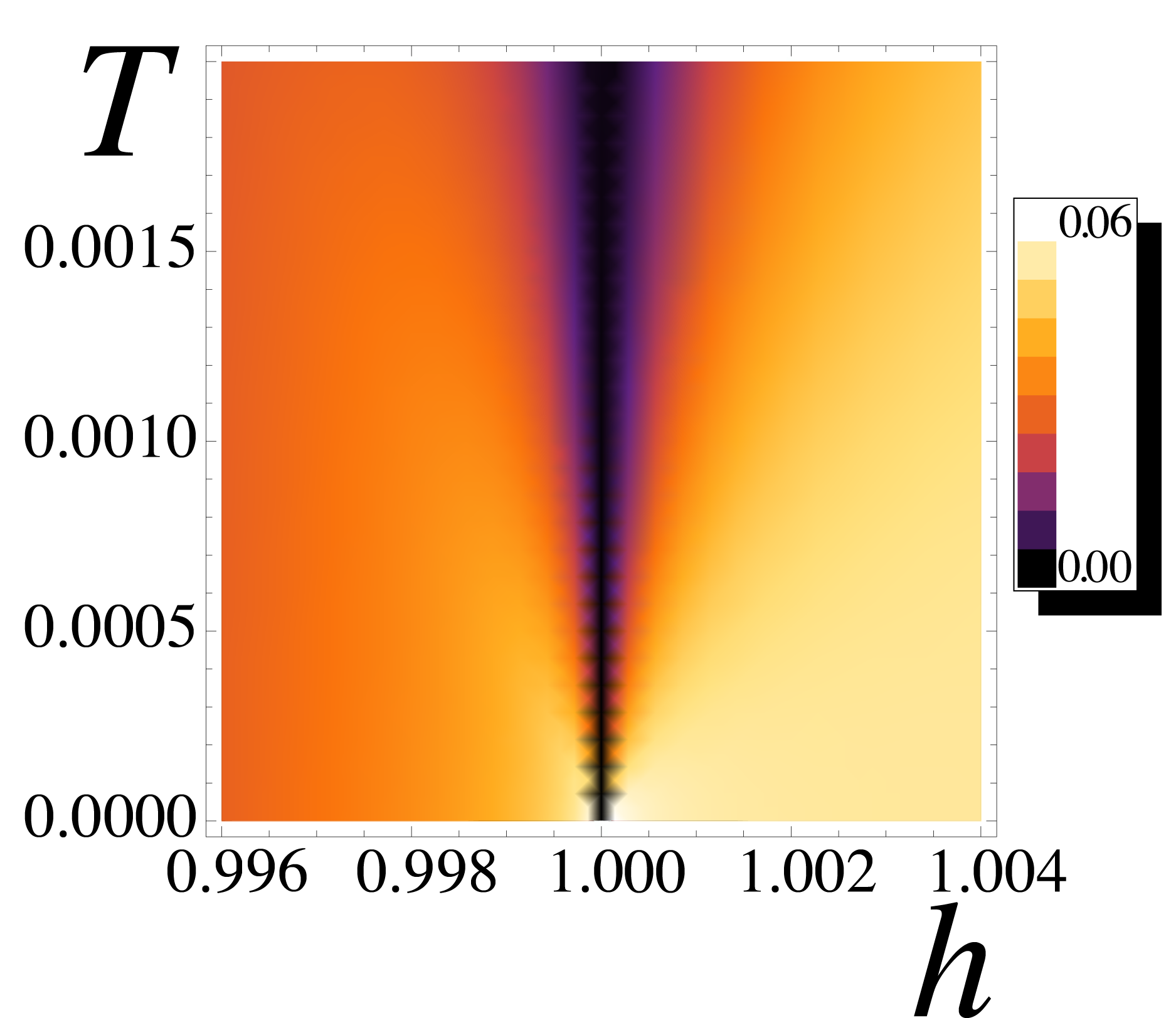}
  \caption{Density plot of $D_{\mathbf{u}}Q_1 $ close to $h_c$, in the $h-T$ plane. 
    The vertical line fanning out from the QCP shows that $ Q_1 $ tends to be constant 
    inside the quantum critical region.}
  \label{deriv_dir}
\end{figure}

\begin{figure}[!t]
  \centering
  \includegraphics[width=0.8\columnwidth]{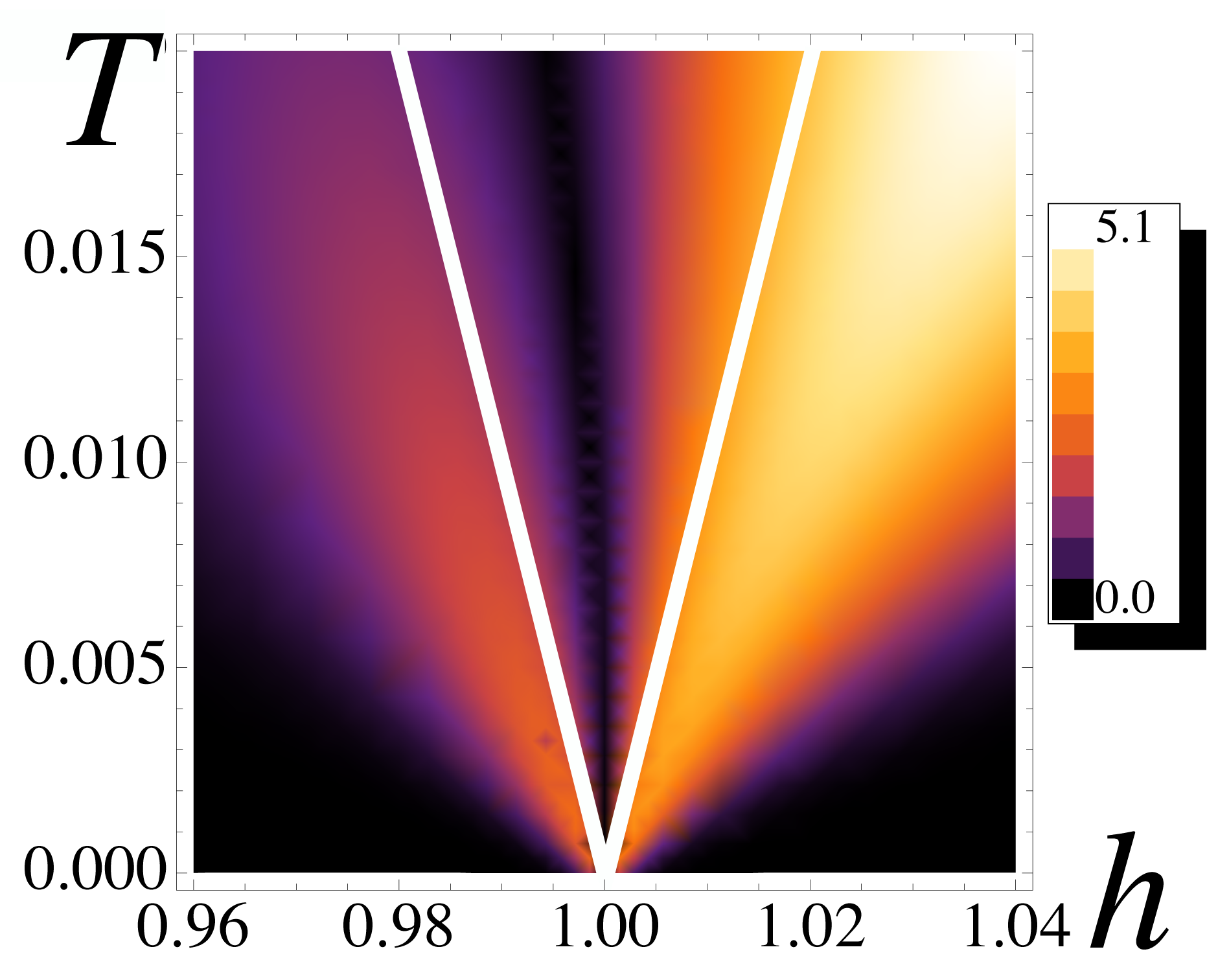}
  \caption{Density plot in the $h-T$ plane of $\partial_T [Q_1/C_1]$ close to $h_c$; 
    along the critical line the ratio $Q_1/C_1$ is constant with respect to the temperature. 
    The solid straight line ( $T= |h-h_c|$ ) marks the boundary of the quantum critical region.}
  \label{ratio_thermal}
\end{figure}

Let us now analyze how criticality and factorization modify the fabric 
of pure quantum correlations in the $h - T$ plane.
We start by focusing on the finite-temperature scaling of the quantum discord close to the critical point $ h_c $.
In the first place we verified the logarithmic scaling $ \partial_h Q_r \vert_{h_c} \sim x \ln(T) + k $
along the critical line, $ h=1 $, in the $h - T$ plane (see Fig.~\ref{ratio_thermal}),
where the value of $ x $ depends on the degree of anisotropy $ \gamma $.
Once $ x $ is given (for example we found $ x= 0.065$, for $r=1$ and $\gamma=0.7$ -- Fig.~\ref{scaling_T}),
by properly tuning the ratio  $ T/T_{cross} $, where $T_{cross} \equiv |h-h_c| $,
we verified the scaling ansatz 
\begin{equation}
\partial_h Q_r = T^x \, F \bigg( \frac{T}{T_{cross}} \bigg) \, .
\label{thermal_ansatz}
\end{equation}
In particular in  Fig.~\ref{scaling_T} we show how different curves, related to different values of $ T/T_{cross} $,
collapse when approaching the critical point.
Remarkably in the Ising case (inset) the scaling is verified as well, even if 
the derivative $\partial_h Q_1$ is finite at $ h_c $.
%
%
To explore the behavior of correlations along the $h - T$ plane, we studied how 
the quantum discord varies on the phase diagram just above the QCP.
In the first place we analyze how the derivative with respect to field
behaves along the directions fanning out from $ h_c $.
In Fig.~\ref{cartoon}  we sketch a cartoon to describe the directional derivative 
$D_{\mathbf{u}}Q_r = \big| \partial_{T}Q_r \sin\alpha + \partial_{a}Q_r \cos \alpha \big|$
we used to describe how $ Q_{1} $ varies close to the QCP.
%
%
From the pattern of $ D_{\mathbf{u}}Q_1 $ at low temperature (Fig.~\ref{deriv_dir})
we see how the presence of the QPT characterizes the whole phase diagram. 
The black vertical line starting from the QCP highlights the fact that 
the quantum discord remains constant along the critical line $h = 1$:
in a sense, close to such region $h \approx 1$, quantum correlations are particularly ``rigid''.
This explains their robustness up to finite temperatures, 
particularly along slopes within the quantum critical region.
On the other hand, out of the quantum critical region, the variation of $ Q_1 $ is drastically increased. 
We also point out the peculiar asymmetric behavior between the two semiclassical regions
(in the ordered phase $ D_{\mathbf{u}}Q_1 $ is generally higher than in the paramagnetic phase).
%
%
Furthermore, to make a more accurate analysis we look 
at the interplay between quantum discord and classical correlations.
In particular we analyze how the ratio $ Q_1/C_1 $ varies 
with the temperature, exploiting the respective sensitivity 
to thermal fluctuations arising at finite temperature.
Accordingly with the phase diagram related to the QPT, 
a $ V $ shaped pattern comes out (see Fig.~\ref{ratio_thermal}).
In particular along the critical line, inside the quantum critical region, we found 
$\partial_T [Q_1/C_1]=0$. Then apparently the ratio between correlations is constant even though 
the temperature is switched on, as long as the field is tuned at the critical value $ h_c $.
Besides this, the whole crossover region from a phase to another onr
is marked as the highest variation in the nature of correlations in the system. 

\begin{figure}[!t]
  \centering
  \includegraphics[width=0.8\columnwidth]{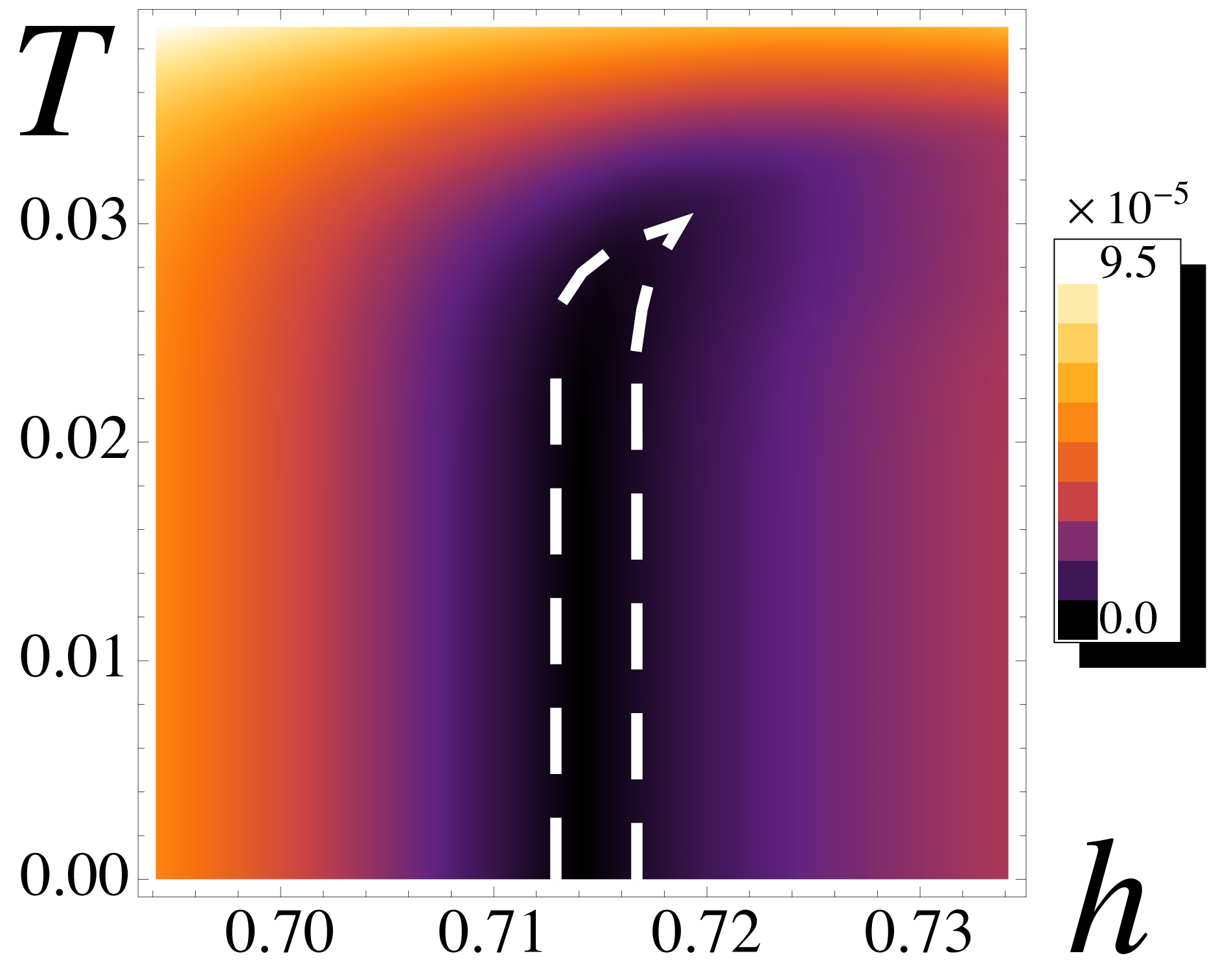}
  \caption{Average quantum discord displacement $\overline{\Delta Q_r}$
    for $m=5$ fanning out from the factorizing point $h_f \sim 0.714$,
    where all correlations coincide at any length scale $r$, as evidenced in the left
    inset of Fig.~\ref{qd}.}
  \label{factorizing_thermal}
\end{figure}

Finally we concentrate on how the factorizing field affects the finite-temperature physics of the system.
As we stressed before, the $ Z_2 $ symmetry is preserved on the thermal state.
In particular, for the thermal ground state, the factorizing field is the unique value 
where the quantum discord is the same at any length scale $ r $.
Here we show that this feature is present even after the temperature is switched on. 
In Fig.~\ref{factorizing_thermal} we propose the quantity
\begin{equation}
  \overline{\Delta Q_r} = \frac{2}{m(m-1)} \sum_{i,j=1}^m |Q_{r_i} - Q_{r_j}|
\end{equation}
as a measure of the robustness of this characteristic at non zero temperature.
We consider different distances between the couple of spins $ A $ and $ B $,
and take the average of the difference between the respective quantum discord. 
Our results strongly support that, for a finite range of temperature, this difference 
is still zero (i.e. the quantum discord for different $ r $ preserves its scale invariance 
only at $ h \approx h_f $).

We emphasize that the robustness and sensitivity of the quantum discord to non zero temperature,
encourage the implementation of suitable experiments that could give good feedback of our analysis.

\section{Discussion}   \label{sec:concl}

We studied pure quantum correlations quantified by the quantum discord $Q_r$ 
in the quantum phases involved in a symmetry-breaking QPT.

In the ordered phase, although $Q_r$ results relatively small in the symmetry-broken state
as compared to the thermal ground state, it underlies key features 
in driving both the order-disorder transition across the QPT at $h_c$, and the 
correlation transition across the factorizing field $h_f$.
The critical point is characterized by a non analyticity of $Q_r$ found in the Ising universality class. 
Close to $h_f$, $Q_r$ displays uniquely non trivial 
properties: in the thermal ground state quantum correlations are identical 
at all scales; for the symmetry-broken state the factorization
can be interpreted as a collective reshuffling of quantum correlations.
We point out that $h_f$ marks the transition between two ``phases'' characterized 
by a different pattern of entanglement~\cite{amico06,fubini}. 
Accordingly our data provide evidence that such a {\it correlation transition} phenomenon 
is of collective nature, governed by an exponential scaling law.
We observe that the scaling close to $h_f$ cannot be algebraic, 
because the correlation functions drop exponentially in the gapped phases.
Nonetheless we ascribe this specific scaling behavior to the peculiar phenomenology 
of the factorizing phenomenon, which goes beyond the generic exponential decay
of correlations that is observed in gapped phases.
For finite $L$ different ground states for the two parity sectors intersect~\cite{giorgidepasquale}.  
The ground state energy density is diverging {\it for all $L$} 
(such divergence, though, vanishes in the thermodynamic limit). 
Indeed we found that the factorization occurs without any violation of adiabatic continuity. 
Accordingly, the ground state fidelity $\mathcal F (h)$,
which can detect both symmetry breaking and non-symmetry breaking QPT, 
is a smooth function at $h_f$~\cite{fidelity}.
We remark that this can occur without closing a gap and changing the symmetry of the system, 
as a signature of the fact that quantum phases and entanglement are more subtle than what 
the symmetry-breaking paradigm says. 
Such a behavior is particularly relevant in the context of QPTs involving topologically 
ordered phases where a QPT consists in the change of the global pattern of entanglement, 
instead of symmetry~\cite{topentropy}. 

We also analyzed the phase diagram at low $ T $. 
A discontinuity of $Q_r$ with $T$ is evidenced in the whole ordered phase $h<h_c$. 
We proved that $Q_r$ displays universal features,
identifying the quantum critical region as the one where the quantum discord 
(relatively to classical correlations) is frozen out to the $T=0$ value. In particular 
in each phase the ratio between correlations is stable with respect to the temperature,
while the highest variations develop in the crossover region 
on the right of $ h_c $, where the system is running out of the critical region into 
the quasi-classical one just above the disordered phase of the paramagnet.
This type of quantum correlations therefore allows a fine structure of 
the phase diagram, according to the behavior of the gap $\Delta \lessgtr 0$ 
in the low temperature limit $ T\ll | \Delta | $:
as a matter of fact, two different mechanisms lead to the two corresponding semiclassical 
regimes driven by quantum ($\Delta > 0$) or thermal ($\Delta< 0$) fluctuations~\cite{sachdev}.  
Finally, a non trivial pattern of quantum correlations fans out from the factorization
of the ground state, as well: we identified a finite portion of the low-temperature 
phase diagram where the quantum discord is nearly constant at any range, 
despite of the thermal fluctuations.

\acknowledgments

We thank A. De Pasquale, R. Fazio, S. Montangero, D. Patan\'e, M. Zannetti, J. Quintanilla for useful discussions.
The DMRG code released within the PwP project (www.dmrg.it) has been used. 
Research at Perimeter Institute for Theoretical Physics is supported in part by the Government of Canada 
through NSERC and by the Province of Ontario through MRI.
DR acknowledges support from EU through the project SOLID.



\begin{thebibliography}{99}

\bibitem{wenbook} 
  X.-G.~Wen,  {\em Quantum Field Theory of Many-Body Systems} (Oxford University Press, USA, 2004).
  
\bibitem{zurek}
  L.~Henderson and V.~Vedral,  J. Phys. A: Math. Gen. {\bf 34}, 6899 (2001);
  H.~Ollivier and W.~H.~Zurek,  \prl {\bf 88}, 017901 (2002).

\bibitem{vedral}
  V.~Vedral,  \prl {\bf 90}, 050401 (2003); 
  B.~Daki\'c, V.~Vedral, and C.~Bruckner,  \prl {\bf 105}, 190502 (2010); 
  A.~Auyuanet and L.~Davidovic,  \pra {\bf 82}, 032112 (2010).

\bibitem{amico} 
  L.~Amico, R.~Fazio, A.~Osterloh, and V.~Vedral,  \rmp {\bf 80}, 517 (2008); 
  L.~Amico and R.~Fazio,  J. Phys. A {\bf 42}, 504001 (2009).

\bibitem{first_ising}
  R.~Dillenschneider,  \prb {\bf 78}, 224413 (2008).

\bibitem{sarandy} 
  M.~S.~Sarandy,  \pra {\bf 80}, 022108 (2009).
 
\bibitem{topo} 
  Y.-X.~Chen and S.-W.~Li,  \pra {\bf 81}, 032120 (2010).

\bibitem{maziero}
  J.~Maziero, H.~C.~Guzman, L.~C.~Celeri, M.~S.~Sarandy, and R.~M.~Serra,  \pra {\bf 82}, 012106 (2010). 

\bibitem{rigolin_temp}
  T.~Werlang, C.~Trippe, G.~A.~P.~Ribeiro, and G.~Rigolin, \prl {\bf 105}, 095702 (2010);
  T.~Werlang, G.~A.~P.~Ribeiro, and G.~Rigolin,  \pra {\bf 83}, 062334 (2011).

\bibitem{recent} 
  B.-Q.~Liu, B.~Shao, J.-G.~Li, J.~Zou, and L.-A.~Wu,  \pra {\bf 83}, 052112 (2011); 
  M.~A.~Yurishchev,  \prb {\bf 84}, 024418 (2011); 
  Bo Li; Y.-S.~Wang, Physica B {\bf 407}, 77 (2012); 
  P.~R.~Wells Jr., B.~Koiller, arXiv:1111.4513. 

\bibitem{sachdev} 
  S.~Sachdev,  {\em Quantum phase transitions} (Cambridge University Press, Cambridge, 2001).

\bibitem{coleman}
  S.~Coleman, ``Secret symmetry: an introduction to spontaneous symmetry breakdown and gauge fields'', 
  in {\em Laws of hadronic matter}, ed. A.~Zichichi (Academic Press, New York, 1975).

\bibitem{factorization} 
  J.~Kurmann, H.~Thomas, and G.~M\"uller,  Physica A {\bf 112}, 235 (1982); 
  T.~Roscilde, P.~Verrucchi, A.~Fubini, S.~Haas, and V.~Tognetti,  \prl {\bf 94}, 147208 (2005).

\bibitem{giampaolo}
  S.~M.~Giampaolo, G.~Adesso, and F.~Illuminati, \prl {\bf 100}, 197201 (2008);
  \prb {\bf 79}, 224434 (2009); 
  \prl {\bf 104}, 207202 (2010).

\bibitem{fubini} 
  A.~Fubini, T.~Roscilde, V.~Tognetti, M.~Tusa, and P.~Verrucchi,  Eur. Phys. J. D {\bf 38}, 563 (2006).
  
\bibitem{tomasello} 
  This paper comes as an extension of the article by
  B.~Tomasello, D.~Rossini, A.~Hamma, and L.~Amico,  Europhys. Lett. {\bf 96}, 27002 (2011).
  Namely we discuss and elucidate, with more details, the results therein contained.

\bibitem{lieb}
  E.~Lieb, T.~Schultz and D.~Mattis,  Ann. Phys. {\bf 16}, 407 (1961).

\bibitem{mcoy}
  E.~Barouch and B.~M.~McCoy,  \pra {\bf 2}, 1075 (1970);  \pra {\bf 3}, 786 (1971).

\bibitem{pfeuty}
  P.~Pfeuty,  Ann. Phys. {\bf 57}, 79 (1970).

\bibitem{amico06} 
  L.~Amico, F.~Baroni, A.~Fubini, D.~Patan\'e, V.~Tognetti, and P.~Verrucchi,  \pra {\bf 74}, 022322 (2006).

\bibitem{palacios} 
  O.~F.~Sylju\aa sen,  \pra {\bf 68}, 060301(R) (2003); 
  A.~Osterloh, G.~Palacios, and S.~Montangero,  \prl {\bf 97}, 257201 (2006);
  T.~R.~de Oliveira, G.~Rigolin, M.~C.~de Oliveira, and E.~Miranda,  \pra {\bf 77}, 032325 (2008).

\bibitem{johnson} 
  J.~D.~Johnson and B.~M.~McCoy,  \pra {\bf 4}, 2314 (1971).

\bibitem{dmrg}
  U.~Schollw\"ock,  \rmp {\bf 77}, 259 (2005).

\bibitem{luo} 
  S.~Luo,  \pra {\bf 77}, 042303 (2008).

\bibitem{ciliberti}
  L.~Ciliberti, R.~Rossignoli, and N.~Canosa,  \pra {\bf 82}, 042316 (2010).

\bibitem{baroni}
  F.~Baroni, A.~Fubini, V.~Tognetti, and P.~Verrucchi,  J. Phys. A: Math. Theor. {\bf 40}, 9845 (2007).

\bibitem{toth} 
  G.~Toth, private communication.

\bibitem{giorgidepasquale} 
  G.~Giorgi,  Phys. Rev. B {\bf 79}, 060405(R) (2009); 
  A.~De Pasquale and P.~Facchi,  Phys. Rev. A {\bf 80}, 032102 (2009).

\bibitem{fidelity} 
  D.~Abasto, A.~Hamma, and P.~Zanardi,  \pra {\bf 78}, 010301(R) (2008).

\bibitem{topentropy} 
  A.~Hamma, R.~Ionicioiu, and P.~Zanardi,  \pra {\bf 71}, 022315 (2005); 
  A.~Kitaev and J.~Preskill,  \prl {\bf 96}, 110404 (2006); 
  M.~Levin and X.-G.~Wen,  \prl {\bf 96}, 110405 (2006).

\end{thebibliography}
\end{document}